# HYDROGEN-RELATED CONVERSION PROCESSES OF GE-RELATED POINT DEFECTS IN SILICA TRIGGERED BY UV LASER IRRADIATION.

F. Messina, M. Cannas

*Dipartimento di Scienze Fisiche ed Astronomiche, Università di Palermo,*

*via Archirafi 36, I-90123, Palermo, Italy*

**Abstract:**

The conversion processes of Ge-related point defects triggered in amorphous $SiO_2$ by 4.7eV laser exposure were investigated. Our study has focused on the interplay between the $(=Ge^{\bullet}\text{-H})$ H(II) center and the twofold coordinated Ge defect $(=Ge^{\bullet\bullet})$. The former is generated in the post-irradiation stage, while the latter decays both during and after exposure. The post-irradiation decay kinetics of $=Ge^{\bullet\bullet}$ is isolated and found to be anti-correlated to the growth of H(II), at least at short times. From this finding it is suggested that both processes are due to trapping of radiolytic $H_0$ at the diamagnetic defect site. Furthermore, the anti-correlated behavior is preserved also under repeated irradiation: light at 4.7eV destroys the already formed H(II) centers and restore their precursors $=Ge^{\bullet\bullet}$. This process leads to repeatability of the post-irradiation kinetics of the two species after multiple laser exposures. A comprehensive scheme of chemical reactions explaining the observed post-irradiation processes is proposed and tested against experimental data.

**PACS**: 71.55Jv; 61.72Cc; 61.80Ba; 82.50Hp

*Corresponding author: F. Messina

Dip.to di Scienze Fisiche ed Astronomiche, via Archirafi 36, I-90123 Palermo.

Phone: +39 0916234218, Fax: +390 916234281, e-mail: fmessina@fisica.unipa.it

# 1. Introduction

The effect of radiation on amorphous silicon dioxide (silica) is a timely research field due to the wide use of $SiO_2$ in advanced optical and electronic technologies and to the ability of radiation to induce stable alterations of the material, often related to generation and conversion processes of point defects.[1-2] In particular, radiation from high-intensity pulsed lasers is mostly effective in inducing *ad hoc* variations of macroscopic properties of $SiO_2$ such as the refraction index; moreover, from a fundamental point of view, the selectivity of laser radiation on precursors allows to conduct comprehensive studies of specific point defect generation and conversion processes.[1-5]

Extrinsic defects due to germanium and hydrogen often play an important role in this effects. In fact, UV exposure of Ge-doped $SiO_2$ causes the generation of Ge-related paramagnetic centers from diamagnetic precursors, this process being considered one of the main causes of photosensitivity of the material.[3,6-10] Experiments have showed that the main diamagnetic precursors defects responding to UV radiation in Ge-containing $SiO_2$ are the fourfold coordinated Ge centers, the oxygen vacancy on threefold coordinated Ge, and the twofold coordinated Ge (=Ge$^{\bullet\bullet}$), also known as germanium lone pair center (GLPC). The GLPC is responsible of absorption peaking at 5.16 eV and emissions at 3.1 eV and 4.2 eV whereas the oxygen vacancy absorbs at 5.06eV not showing any measurable emission.[6-12] Hydrogen, being mobile in the amorphous matrix even at room temperature, takes part to diffusion-limited reactions with induced or pre-existing point defects, altering their concentration also in the post-irradiation stage and so influencing the response of silica to radiation.[13-17] The photochemical transformation mechanism of Ge-related defects as well as $H_2$-related effects currently remain an open topic of investigation of $SiO_2$, since many issues are not yet thoroughly understood.

Silica obtained by fusion of natural quartz powder (natural silica) usually contains a small concentration (≈1ppm) of Ge impurities due to natural contamination, which are mainly arranged in the GLPC form.[18] In recent studies it has been observed that in natural silica the main Ge-related paramagnetic defect induced by fourth harmonic Nd:YAG (4.7eV) laser exposure at room temperature is the H(II) center (=Ge$^\bullet$-H), common also in irradiated H$_2$-loaded Ge-doped SiO$_2$ and under γ irradiation, and detectable by electron spin resonance (ESR).[9,17,19-21] Also Ge(2) centers (=Ge$^\bullet$) are UV-induced, but in much lower concentration than H(II).[22] The growth of H(II) occurs mainly in the post-irradiation stage, when UV exposure is over, and was suggested to occur by trapping of diffusing H$_0$ at the GLPC site:[22]

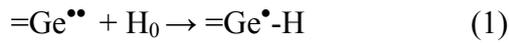

$$=Ge^{\bullet\bullet} + H_0 \rightarrow =Ge^\bullet\text{-H} \qquad (1)$$

in this scheme, H$_0$ required at the left side of reaction (1) is made available by breaking of H$_2$ on the paramagnetic E' (≡Si•),[23] induced as well by laser irradiation:

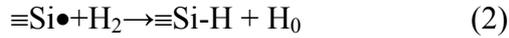

$$\equiv Si\bullet + H_2 \rightarrow \equiv Si\text{-H} + H_0 \qquad (2)$$

consistently, E' are observed to decrease in the post-irradiation stage.[24] Hydrogen is of photolytic origin, produced from Si-H or O-H groups and subsequent dimerization.[25]

The attribution of the post-irradiation kinetics of E' and H(II) to diffusion and reaction of H$_2$ was made on the basis of a semi-quantitative comparison of the typical time scale of the processes with the diffusion parameters of the mobile specie.[17,24-25] However it is worth to note that the observation by ESR of the hydrogen-related H(II), whose concentration increases in time, naturally leads to attribute the post-irradiation effects to H$_2$, ruling out *a priori* other possibilities, like electron/hole detrapping or diffusion of other mobile species. Then, H(II) may be considered a probe of the presence of mobile hydrogen.

However, many issues regarding H(II) and its relationship with GLPC remain open: the correlation between bleaching of GLPC and H(II) formation due to reaction (1) still has to be analyzed; furthermore, the overall compatibility of reactions (1) and (2) with the detailed time dependences of H(II), GLPC and E' must be demonstrated.

We present here a study of the conversion processes of Ge-related defects in silica, elicited by exposure of the material to 4.7eV laser light. Our main purpose is to clarify the interplay between GLPC and H(II) defects triggered by UV exposure, including in our analysis also the effect of repeated irradiations on a single sample. More in general, these experiments aim to develop a comprehensive interpretation of post-irradiation effects characteristic of these materials and related to diffusing mobile hydrogen.

2. Materials and methods

As in natural silica Ge is mainly arranged in twofold coordinated form,[18,22] this is a material of choice to study selectively the conversion processes of GLPC, since it almost lacks of three- and four-fold coordinated Ge precursors, common in heavy Ge-doped $SiO_2$.[6-8,18] For this reason, two commercial natural silica types were used in our experiments: a type I dry EQ906 supplied by Quartz & Silice, OH content ~20ppm, and a type II wet HERASIL1, OH content ~150ppm, supplied by Heraeus QuartzGlas. These specimens (5x5x1$mm^3$ shaped) are obtained by fusion of natural α-quartz powder by electric arc in an inert gas atmosphere (dry) or by a $H_2/O_2$ flame (wet). Samples contain Ge impurities in $(1.5\pm0.3)\times10^{16}$ $cm^{-3}$ concentration, as determined with the neutron activation technique.[18] Consistently with the properties of Ge impurities in natural $SiO_2$, previous studies have showed that the native absorption band at ~5eV ($B_{2\beta}$ band) detected in as-grown samples is due exclusively to twofold coordinated Ge (and not to oxygen vacancies) and is linearly correlated with the 3.1eV and 4.2eV emissions[26]. The peak amplitude of $B_{2\beta}$ in our samples prior to irradiation was measured to be $(0.43\pm0.04)$ $cm^{-1}$ and $(0.28\pm0.03)$ $cm^{-1}$ in EQ906 and HERASIL1 respectively.

UV exposure with 4.7eV photons was performed at room temperature using the fourth harmonic from the pulsed radiation of a Quanta System SYL 201 Nd:YAG laser, at a repetition rate of 1 Hz, each fourth-harmonic pulse having energy density of W=40mJ/$cm^2$ and 5ns duration.

H(II) center was detected at room temperature by ESR measurements on its characteristical 11.8mT hyperfine doublet, due to the interaction between the unpaired electronic spin on Ge and the nuclear spin of the proton H.[21] The signal was detected on a spectrometer (Bruker EMX) working at 9.7GHz with microwave power P=3.2mW, small enough to prevent saturation, and a 100 kHz modulation field of peak-to-peak amplitude, $B_m$=0.4mT. The uncertainty on EPR signal intensity is 10%. The absolute concentration of the paramagnetic centers was calculated by comparing the double-integrated ESR spectra with that of E' centers, whose absolute density was determined with accuracy of ±20% by spin-echo measurements.[27]

Optical absorption (OA) spectra in UV range were acquired by a JASCO V-570 double beam spectrophotometer, with a $D_2$ lamp source and using a 2 nm bandwidth.

Photoluminescence (PL) measures were carried out with a JASCO FP-770 spectrofluorometer with a 150W Xe-lamp source; all PL spectra reported in this work were obtained with a 3 nm excitation and a 3 nm emission bandwidths and were corrected for spectral sensitivity and dispersion of detecting system. Luminescence emission spectra of GLPC were measured under 5.0eV excitation, falling well into its $B_{2\beta}$ absorption band.[2,11] Being the optical density of our samples at 5.0eV smaller than 0.02, the luminescence intensity can be considered to be proportional to the concentration of the GLPC center.[26]

To follow the PL intensity variation in the post-irradiation stage, ΔPL, the irradiated specimens were positioned into the sample chamber of the spectrofluorometer about $10^2$s after exposure, after which they were kept in place and measured for $10^4 \div 10^5$s; with this choice, the precision of ΔPL is increased and estimated to be 2% of PL intensity.

## 3. Results and Discussion

### A. Conversion Mechanisms

In Figure 1 is shown the kinetics of H(II) centers observed in HERASIL1 and Q906 samples after the end of a 2000 pulses laser irradiation. The concentration [H(II)] of the paramagnetic defects was measured from the intensity of their 11.8mT ESR doublet (shown in the inset) at different delays after the end of UV exposure. Results are shown in Fig. 1 (a) for the wet and Fig. 1 (b) for the dry specimen. Here and in the other graphs, the origin of the time scale corresponds to the end of exposure. The main evidence is the post-irradiation growth of [H(II)], which increases from the initial values of $(3.7\pm0.4)\times10^{14}cm^{-3}$ (wet) and $(6.0\pm0.6)\times10^{14}cm^{-3}$ (dry), measured at $t\sim10^2$s, to the stationary values of $(1.9\pm0.2)\times10^{15}cm^{-3}$ (wet) and $(1.4\pm0.1)\times10^{15}cm^{-3}$ (dry), about $10^5$s after the end of illumination.

We irradiated in the same conditions another HERASIL1 and another Q906 sample, to find out if the GLPC undergoes a post-irradiation kinetics concurrent to the growth of H(II). To this aim, PL measurements under lamp excitation at 5.0eV were performed on the two samples, before irradiation and then at different delays ($10^2\div10^5$s) from the end of exposure. The detected emission spectra consist in the 3.1eV and 4.2eV bands, whose UV excitation spectrum closely resembles the $B_{2\beta}$ band. The overall spectroscopic picture is consistent with our attribution of this PL activity to the GLPC center[11,26].

We found that irradiation induces a bleaching of the emission bands occurring in two clearly distinguishable stages: (a) during illumination, an intensity reduction of ~50% in HERASIL1 and ~15% in Q906 takes place, as we observe by comparing the as-grown PL spectrum (not reported) with the first detected after exposure (at $t\sim10^2$) (b) after the end of irradiation, the PL intensity further decreases in time as evidenced by the spectra in Fig. 2, measured in the HERASIL1 sample at different delays ($10^2\div10^5$s) from the end of exposure. An analogous result was obtained on the Q906 specimen. In both cases, measures were continued until a constant PL intensity was reached within experimental error.

The post-irradiation kinetics of the GLPC is summarized in Figure 3, where the integrated intensity PL(GLPC) of the signal is plotted for the two materials against time. In HERASIL1, panel (a), luminescence intensity decreases of 0.40±0.03 a.u. from 65s to $8\times10^4$s. In Q906, panel (b), the decrease is 0.33±0.07 a.u from 60s to $6\times10^3$s.

To deeper analyze the relationship between H(II) and GLPC, we investigated the concentration variations of both defects under repeated irradiations. In detail, an experiment was performed in which a HERASIL1 specimen was irradiated 3 times with 2000 laser pulses; after each exposure, the post-irradiation kinetics of PL(GLPC) was measured until completion. Results are shown in Fig. 4, panel (a). On a second sample subjected to the same irradiation sequence, the post-irradiation kinetics of H(II) centers was measured after each exposure (panel (b) of Fig. 4).

As apparent from experimental data, each exposure destroys most of H(II) which had formed upon the previous illumination, their concentration decreasing approximately to the same value as immediately after the previous irradiation; simultaneously, we observe a rebuild of luminescence intensity to approximately the same value found at the same time after the previous exposure. After every re-irradiation, the sample loses memory of its previous history, meaning that both PL(GLPC) and [H(II)] repeat again the same decrease/growth kinetics. We stress that the repeatable decay and recovery cycles of GLPC observed upon multiple irradiations involve only the portion bleached in the post-irradiation stage, whereas the reduction observed during exposure occurs irreversibly only during the earliest irradiation.[28]

Since PL(GLPC) is proportional to the concentration of the twofold coordinated center, the bleaching induced by irradiation is a manifestation of conversion processes triggered by UV exposure which transform the diamagnetic center in other defects. The proportionality coefficient ε = [GLPC]/PL(GLPC) between concentration of GLPC and PL intensity (expressed in arbitrary units) can be calculated from (i) the known constant ratio between PL intensity and absorption band area A($B_{2\beta}$),[26] and (ii) the oscillator strength $f_\tau$ of GLPC, estimated from the radiative singlet-singlet decay time, measured at T=10K under synchrotron radiation.[29] With this procedure, we estimate $\varepsilon=(4.4\pm0.7)\times10^{15}$cm$^{-3}$.

The analysis of the time dependence of the GLPC conversion allows to isolate two different stages of the process: that occurring only once during the earliest irradiation, and that taking place after each irradiation. First of all, we briefly comment on the former stage. From $\varepsilon$, we estimate the concentrations of the centers converted during the first irradiation: $\Delta_0=(6.0\pm0.9)\times10^{15}$cm$^{-3}$ (wet) and $\Delta_0=(2.7\pm0.4)\times10^{15}$cm$^{-3}$ (dry). Present data do not permit to clarify in what defect is converted this portion of GLPC, so leaving open this specific issue. Though, we can exclude H(II) and Ge(2), whose concentration at t=0 ([Ge(2)]<$2.5\times10^{14}$cm$^{-3}$) is too small to account for $\Delta_0$,[22] and Ge-E' and Ge(1),[30,31] which are absent in the exposed specimen within the EPR sensitivity of $\sim2\times10^{14}$cm$^{-3}$. Then, we infer that during the earliest irradiation a portion of GLPC is most likely converted in some unknown diamagnetic center which happens to be virtually invisible at this concentration. We point out that this finding contrasts with the common practice in literature to correlate the reduction of GLPC with the concentration of induced paramagnetic signals.[6-9,12]

Since the initial irreversible decay of the diamagnetic defect during the earliest irradiation is not related to generation of H(II), the discussion hereafter will be focused only on what is observed in the post-irradiation stage and upon repeated irradiations. Hence, we proceed to examine the relation between the post-irradiation decay of GLPC and the simultaneous growth of H(II).

Since, for the reasons discussed in the introduction, we consider mobile $H_2$ the cause of the observed post-irradiation processes in natural silica,[17,24-25] we are necessarily led to ascribe the PL decrease of Figs. 2 and 3 to H-trapping at the twofold coordinated Ge site; hence, the post-irradiation kinetics of PL(GLPC) is due to reaction (1) which forms H(II) from the diamagnetic centers.

In this scheme, the decay of GLPC and the growth of H(II) are expected to occur with anti-correlated kinetics. To investigate this issue, in Fig. 5 the increase $\Delta$H(II) of [H(II)] from t$\approx$10$^2$s is plotted for both materials against the decrease in GLPC concentration calculated from the same time instant: $-\Delta$[GLPC]=$-\varepsilon\Delta$PL. EPR data were obtained by extrapolation at the same time instants at which luminescence spectra had been acquired.

We see that data from both materials sit on a single line for short times, whereas for long times they tend to go away from the line towards the upper semiplane. In the short time region, H(II) and GLPC are indeed anti-correlated, with a correlation coefficient independent from the material and represented by the slope of the line, S ~ 0.7, as estimated by a best fit procedure on the first points (corresponding to t<2×10$^3$s). This value of S, which is founded on two completely independent concentration measurements, can be considered to be in good agreement with unity for all present purposes.

These findings suggest the following interpretation of the behaviour of Fig. 5: the linear relationship approximately valid at short times represents a one to one conversion between GLPC and H(II) centers by process (1), whereas the deviations from linear correlation indicate that H(II) centers are generated also by a second channel prevailing on reaction (1) at long times. Since we detect Ge(2) centers in EQ906 samples after irradiation,[22] a possible mechanism producing the portion of H(II) not anti-correlated to GLPC may be the successive H$_0$ and e$^-$-trapping on Ge(2), as recently proposed by Fujimaki et al.[9] In both samples, this second generation channel accounts for ~30% of the total $\Delta$[H(II)].

To further strengthen the relation between H(II) and GLPC, from Fig. 4 we see that the two defects show an anti-correlated behavior also under repeated irradiation. In fact, each exposure causes a photo-decay of the H(II) generated during the last post-irradiation kinetics, and simultaneously restores the GLPC. The concurrence of the two processes, combined with the known structural relationship between the two defects, suggest the following microscopic mechanism responsible for the photo-decay of H(II):

$$=Ge^{\bullet}\text{-H} + h\nu \rightarrow =Ge^{\bullet\bullet} + H_0 \quad (3)$$

the exposure of H(II) to 4.7eV photons causes detaching of the hydrogen atom from the Ge-H bond reconstructing the precursor GLPC. To measure the cross section $\sigma_D$ of process (3), we performed a further experiment starting from a HERASIL1 sample exposed to 2000 laser shots. After waiting the H(II) post-irradiation kinetics to be completed, we measured [H(II)]=$(2.0\pm0.2)\times10^{15}$cm$^{-3}$. Then, we irradiated again the specimen with an increasing number of shots, and between successive exposures we measured the defect concentration. Consistently with Fig. 5, we observed that the new irradiation results in the destruction of H(II) which had grown at after the first dose, their concentration decreasing to ~25% of the initial value after ~250 laser pulses (Fig. 6). An identical effect was observed also on Q906, in agreement with our interpretation in which it is due to the direct absorption of UV light at the defect site (process (3)). The H(II) reduction with the number of pulses N is fitted by an exponential function :

$$[H(II)]=A_1\times\exp(-N/N_0)+A_2 \quad (4)$$

where $N_0=30\pm3$, $A_1=(1.5\pm0.1)\times10^{15}$cm$^{-3}$, $A_2=(0.5\pm0.1)\times10^{15}$cm$^{-3}$. From $N_0$, we calculate: $\sigma_D=N_0^{-1}(h\nu/W)=(6.2\pm0.6)\times10^{-19}$cm$^2$.

Finally, we stress that the observation of the photo-induced decay of H(II) centers allows us to understand the feature of the growth kinetics of these defects, that are formed mostly in the post-irradiation stage, rather than during it, as apparent from data in Fig. 1 (the final concentrations are ~2÷3 times larger than initial concentrations). This result is explained as follows: formation of H(II) during irradiation through reaction (1) is inhibited because of the competition with the photo-induced decay of the centers.[32]

### B. H$_2$ Diffusion-Limited Reaction Kinetics

In the model depicted so far, H(II) are mainly formed by trapping of H$_0$ at the precursor GLPC site, where H$_0$ is made free by breaking of H$_2$ on E' centers. This multi-step process is described by reactions (1) and (2), to which one must add equation (5) accounting for the possibility that $H_0$ produced by reaction (2) is then trapped on another E' center:

$$\equiv Si\bullet + H_0 \rightarrow \equiv Si-H \qquad (5)$$

It is necessary to find out if the model inherent in reactions (1),(2),(5) is capable of describing in detail the measured time dependencies of the three observed species, so testing our attribution of the post-irradiation processes to diffusing hydrogen. To this aim, we start from the chemical rate equations governing the kinetics of (1),(2),(5):

$$\frac{d[E']}{dt} = -k_2[E'][H_2] - k_4[E'][H_0]$$
$$\frac{d[Ge^{\bullet\bullet}]}{dt} = -\frac{d[H(II)]}{dt} = -k_1[Ge^{\bullet\bullet}][H_0]$$
$$\frac{d[H_2]}{dt} = -k_2[E'][H_2] \qquad (6)$$
$$\frac{d[H_0]}{dt} = k_2[E'][H_2] - k_4[E'][H_0] - k_1[Ge^{\bullet\bullet}][H_0]$$

Where $k_1, k_2, k_4$ are the rates of reactions with the same index and are mainly determined by the diffusion coefficients of $H_2$ ($k_2$) and $H_0$ ($k_1, k_4$).

Based on the much higher diffusion constant of $H_0$ with respect to $H_2$, equations (6) can be simplified by the stationary-state approximation, which consists in setting $d[H_0]/dt \approx 0$.[33-34] Then, the last of the (6) can be used to eliminate $[H_0]$ from the other equations, which become:

$$\frac{d[E']}{dt} = -k_2[E'][H_2]\left(1 + \frac{1}{1 + R\frac{[Ge^{\bullet\bullet}]}{[E']}}\right)$$

$$\frac{d[Ge^{\bullet\bullet}]}{dt} = -\frac{d[H(II)]}{dt} = \frac{d[E']}{dt} = -k_2[E'][H_2]\left(1 - \frac{1}{1 + R\frac{[Ge^{\bullet\bullet}]}{[E']}}\right) \qquad (7)$$

$$\frac{d[H_2]}{dt} = -k_2[E'][H_2]$$

$R=k_1/k_4$ is a parameter which controls the ratio between the portions of $[H_0]$ re-captured by E' and the one forming H(II) centers. From (7), the main parameter controlling the overall kinetics is the reaction constant $k_2$ between E' and $H_2$; In Waite's model of diffusion-limited reactions, $k_2$ can be written as $k_2=4\pi r_0 D_0 \exp(-E_a/kT)$, where $r_0$ is the capture radius for the reaction between E' and $H_2$, expected to be of the order of $10^{-8}$cm,[33] while $E_a$ and $D_0$ are pre-exponential factor and activation energy for diffusion of mobile molecular hydrogen, reported to be $E_{aS}=0.45$eV and $D_{0S}=5.65\times10^{-4}$cm$^2$s$^{-1}$.[35, 36]

We measured the post-irradiation kinetics of [E'] in a HERASIL1 sample subjected to 2000 laser pulses from the amplitude of the 5.8eV OA band and using the known value of the peak absorption cross section of the paramagnetic center;[2] results are reported in Fig. 7 as full square points. In the same graph are reported again the kinetics of [H(II)] (from Fig. 1), and of [GLPC], calculated from data in Fig. 3 using the conversion coefficient ε. Then the solutions of system (7), found numerically, were fitted to the experimental datasets.

In the fitting procedure, the initial concentrations of E', GLPC and H(II) were constrained to the values obtained by extrapolating the experimental curves at t=0: $[E'](t=0)=(8.6\pm0.5)\times10^{15}$cm$^{-3}$; $[GLPC](t=0)=(4.7\pm0.2)\times10^{15}$cm$^{-3}$; $[H(II)](t=0)=(2.0\pm0.2)\times10^{14}$cm$^{-3}$. Hence, the fitting parameters which remain to be determined are $[H_2](t=0)$, R and $k_2$; the best fit values of the first two were found to be $[H_2](t=0)=(4.1\pm0.3)\times10^{15}$cm$^{-3}$; R = 1.1±0.2. For what concerns $k_2$, a more complex picture emerges. In fact, as already known from literature,[13,16] we found that a good fit to the data on all the $10^2 \div 10^6$s time scale can be achieved only by using a linear combination of solutions of (7) obtained for different values of $E_a$. This is commonly interpreted as a consequence of the amorphous nature of silica, which manifests itself in a statistical distribution of diffusion activation energies. In detail, dotted lines in Fig. 7 represent typical solutions of system (7) obtained with a single value of $E_a$, which manifestly fail to reproduce the shape of the experimentally observed kinetics. At variance, we found that an excellent agreement (solid curves in Fig. (7)) is attained introducing a gaussian distribution of $E_a$ with mean $<E_a>=0.55\pm0.01$eV and FWMH $\Delta E_a=0.12\pm0.02$eV, where the pre-exponential factor was set to $D_{0S}$, whereas the capture radius was arbitrarily chosen to be $r_0=10^{-8}$cm.

The finding that all three independent experimental datasets can be fitted *at once* for a suitable choice of parameters is a clear proof of the validity of the chemical model hypothesised to explain the post-irradiation processes. This is particularly true if we consider the approximations implicit in equations (6), such as having neglected every other generation channel of H(II) and other possible reactions like the recombination of two $H_0$.

A few comments on the values of the fitting parameters:

a) $<E_a>=0.55$ is higher than the $E_{aS}=0.45$eV value reported for $H_2$ diffusion in $SiO_2$.[13,36] This probably means that the passivation of E' by $H_2$ is also reaction-limited, i.e. the value of $<E_a>$ incorporates also the activation energy for reaction. Actually, present data do not allow to separate the contribution of $r_0$ and $D_0$ to the reaction constant $k_2$, so that the distribution of $E_a$ could be shifted with a different choice of $r_0$ or $D_0$. In fact, only $\Delta E_a$ has an absolute meaning and is in good agreement with previously found distribution widths.[13,16]

b) At t=0, concentration of $H_2$ is approximately one half of E', this suggesting a specific interpretation: E' and $H_0$ are generated during irradiation from the common precursor Si-H, as already proposed elsewhere.[37] Finally, we note that a value of R of the order of unity is to be expected in the framework of Waite's model: in fact, R should equal the ratio of the capture radiuses of GLPC and E' for $H_0$, which both should be of the order of an atomic dimension if the diffusion-limited approach is applicable at all.

In principle, the post-irradiation kinetics are determined by the initial concentration of all defects at the end of laser exposure. For this reason, the observed repeatability (Fig. 5) of the post-irradiation kinetics of H(II) and GLPC after re-irradiation implies that a 2000 pulses exposure has the ability to reset the defects to fixed concentration values independent from the previous history of the sample. Indeed, a similar memory loss effect has consistently been observed also for E' centers in wet natural $SiO_2$,[24] which are involved in the reactions as well, playing the role of $H_2$-cracking centers. As regards H(II), a necessary step to achieve memory loss is the photochemical decomposition (3), which permits to temporarily destroy the paramagnetic center still recovering its precursor, which is then available for the next post-irradiation kinetics.

## 5. Conclusions

The conversion processes of Ge-related defects in silica irradiated by 4.7eV laser light were investigated. We observed the post-irradiation growth of H(II) center as well as the decay of the PL activity associated to GLPC center. The analysis of the time dependence of GLPC signal permits to isolate the post-irradiation stage of its conversion process, which is ascribed to trapping of $H_0$ at the defect site leading to the generation of H(II) center, on the basis of the anti-correlated concentration variations of the two species. This process can be reversed by a second laser exposure, which destroys H(II) and restores the precursor GLPC. Due to this mechanism, the material loses memory upon re-irradiation, meaning that the two centers repeat the same post-irradiation kinetics after multiple exposures. Atomic hydrogen to be trapped on the diamagnetic precursor is produced by breaking of diffusing $H_2$ on E' centers. Consistently, the time dependence of E', H(II), and GLPC concentrations can be fitted by a suitable set of coupled rate equations describing the chemical reactions triggered by irradiation.

## 6. Acknowledgements

We wish to thank our research group at University of Palermo led by Prof. R. Boscaino for support and useful discussions. Technical assistance by G. Lapis and G. Napoli is also acknowledged. This work is part of a national project (PRIN2002) supported by the Italian Ministry of University Research and Technology.

the paramagnetic centers from being destroyed during PL measurements, to acquire all PL measurements reported in this work we used a high wavelength scan speed (500nm min$^{-1}$) in order to reduce as much as possible the exposure time of the sample to the lamp. We verified that this choice, though reducing the signal/noise ratio, avoids the photo-decay of H(II) as well as any appreciable distortion of the experimentally observed PL kinetics.

**FIGURE CAPTIONS**

**FIG.1:** Post-irradiation kinetics of H(II) centers in (a) wet and (b) dry $SiO_2$, as measured from the intensity of the ESR 11.8mT hyperfine doublet of the paramagnetic defect, shown in the inset.

**FIG.2:** PL emission spectra excited at 5.0eV of GLPC center in wet $SiO_2$ after different delays from the end of a 2000 pulses Nd:YAG laser irradiation.

**FIG.3:** Time dependence in the post-irradiation stage of the intensity PL(GLPC) of the PL emission signal associated to GLPC, as observed in (a) wet and (b) dry natural $SiO_2$.

**FIG.4:** Kinetics of (a) PL(GLPC) and (b) [H(II)] induced in a wet $SiO_2$ specimen by a cycle of 4 repeated laser exposures, 2000 shots for each one.

**FIG.5:** Correlation plot between the increase of H(II) concentration and the decrease of GLPC concentration, both measured from $t_0 \approx 10^2$s after the end of irradiation. Full and empty symbols represent respectively wet and dry $SiO_2$

**FIG.6:** Variations of [H(II)] induced by re-irradiation of a natural wet silica sample preliminary exposed to 2000 laser shots; solid line plots the exponential best fit of the data.

**FIG.7:** Concentrations of (squares) E', (circles) H(II), (triangles) GLPC centers after 2000 laser shots in a wet specimen. Dotted lines are obtained by solving a system of rate equations suitable to describe the reactions responsible for the post-irradiation kinetics; solid lines take also into account the statistical distribution of $H_2$ diffusion activation energy $E_a$ typical of amorphous $SiO_2$.

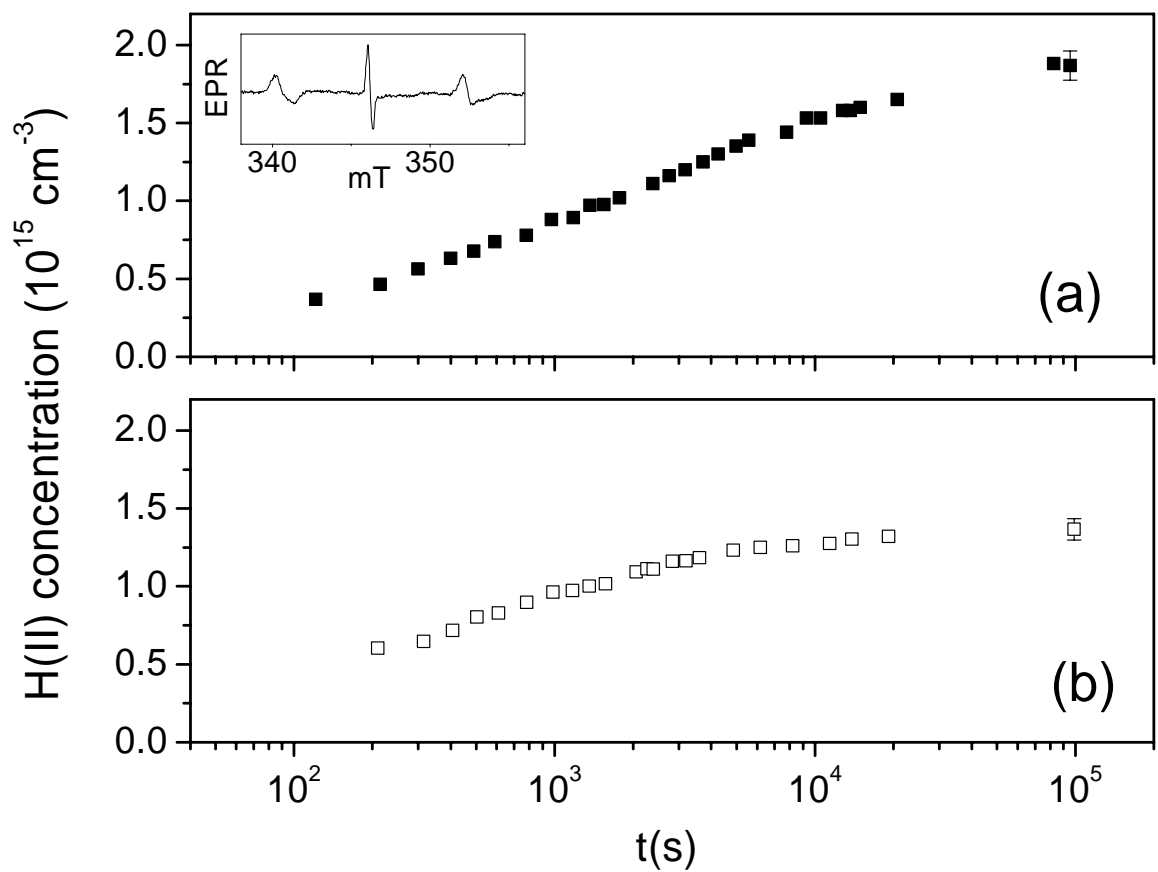

**FIGURE 1**

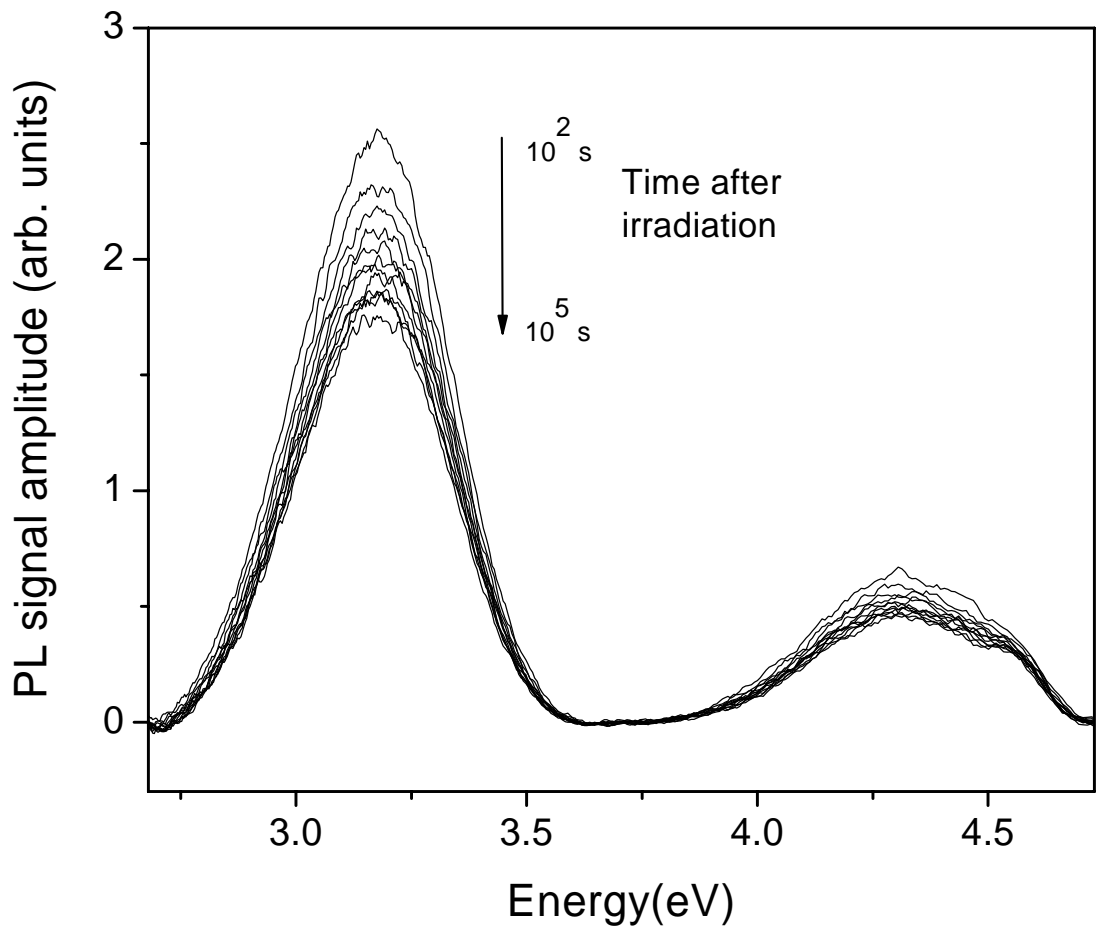

**FIGURE 2**

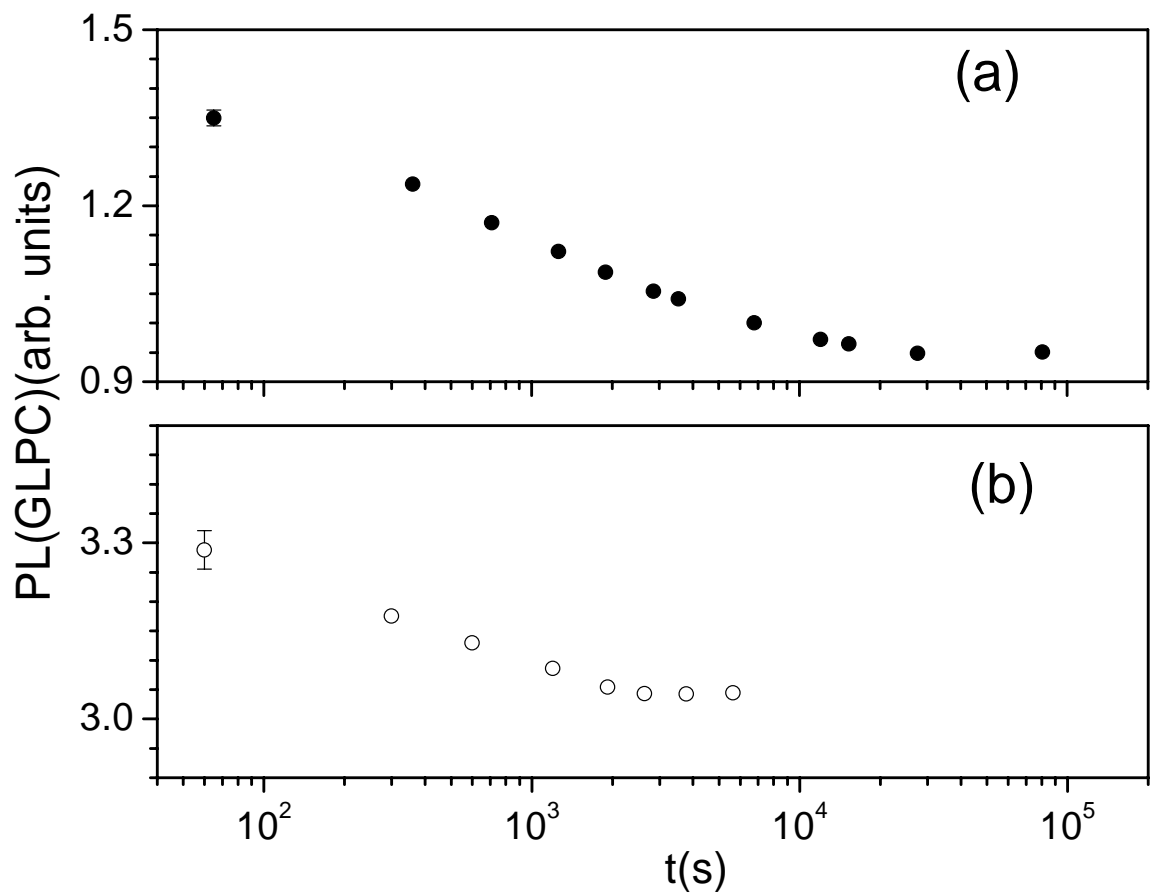

**FIGURE 3**

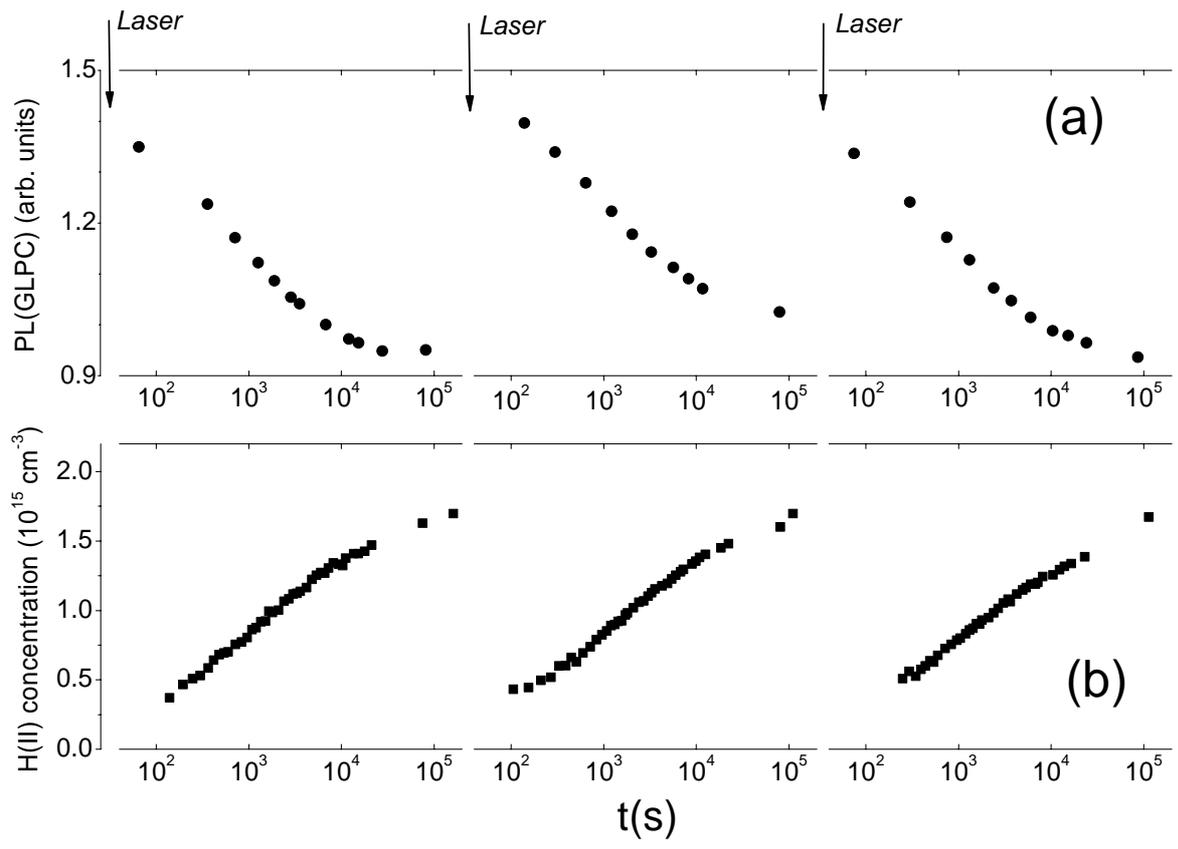

**FIGURE 4**

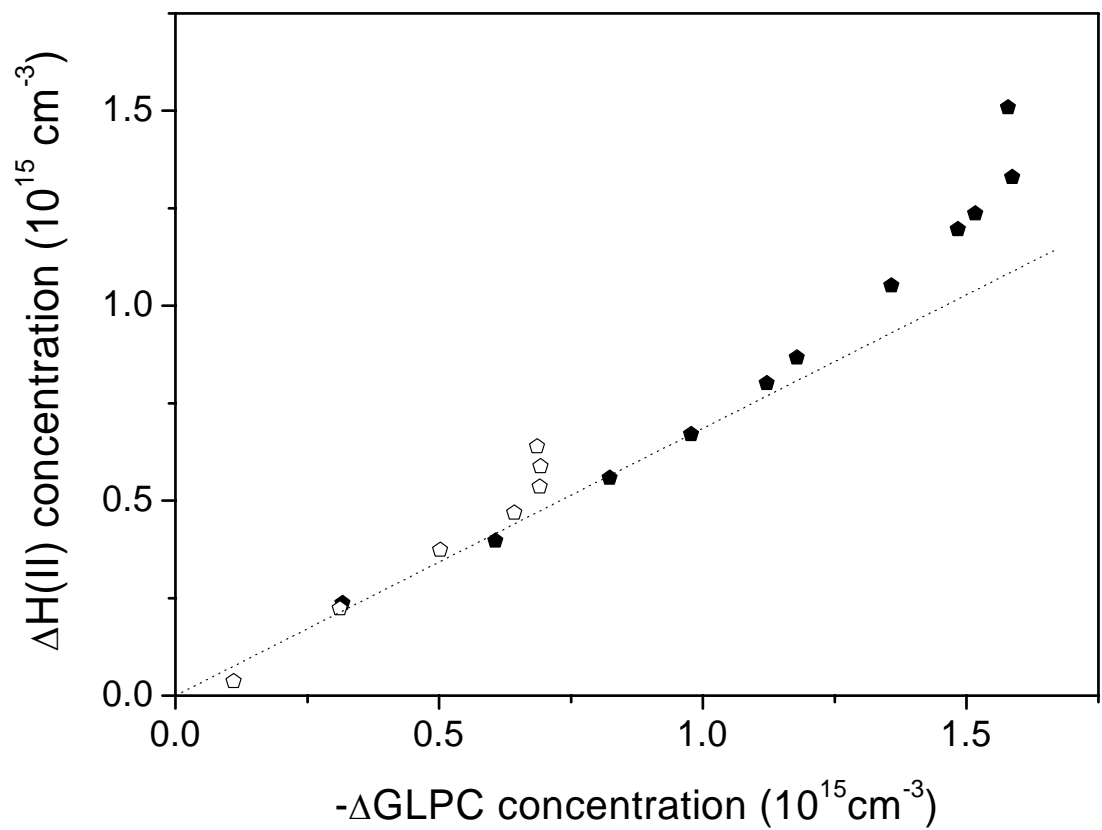

**FIGURE 5**

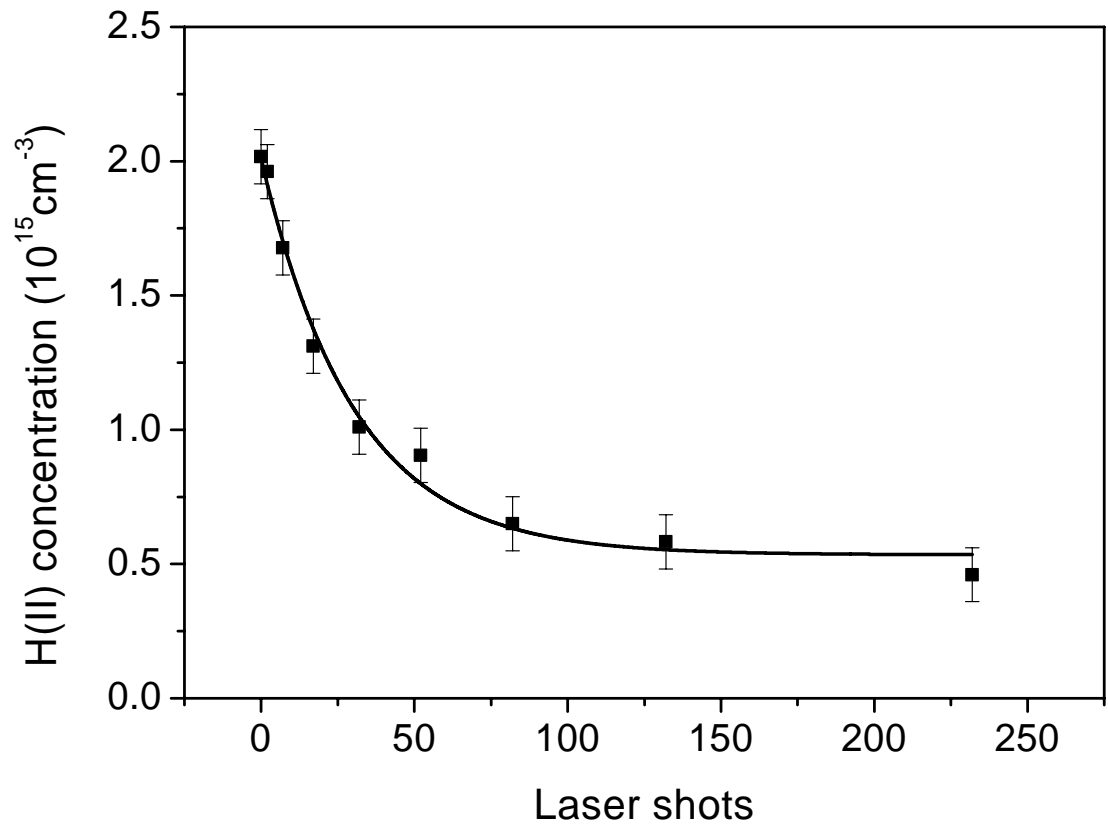

**FIGURE 6**

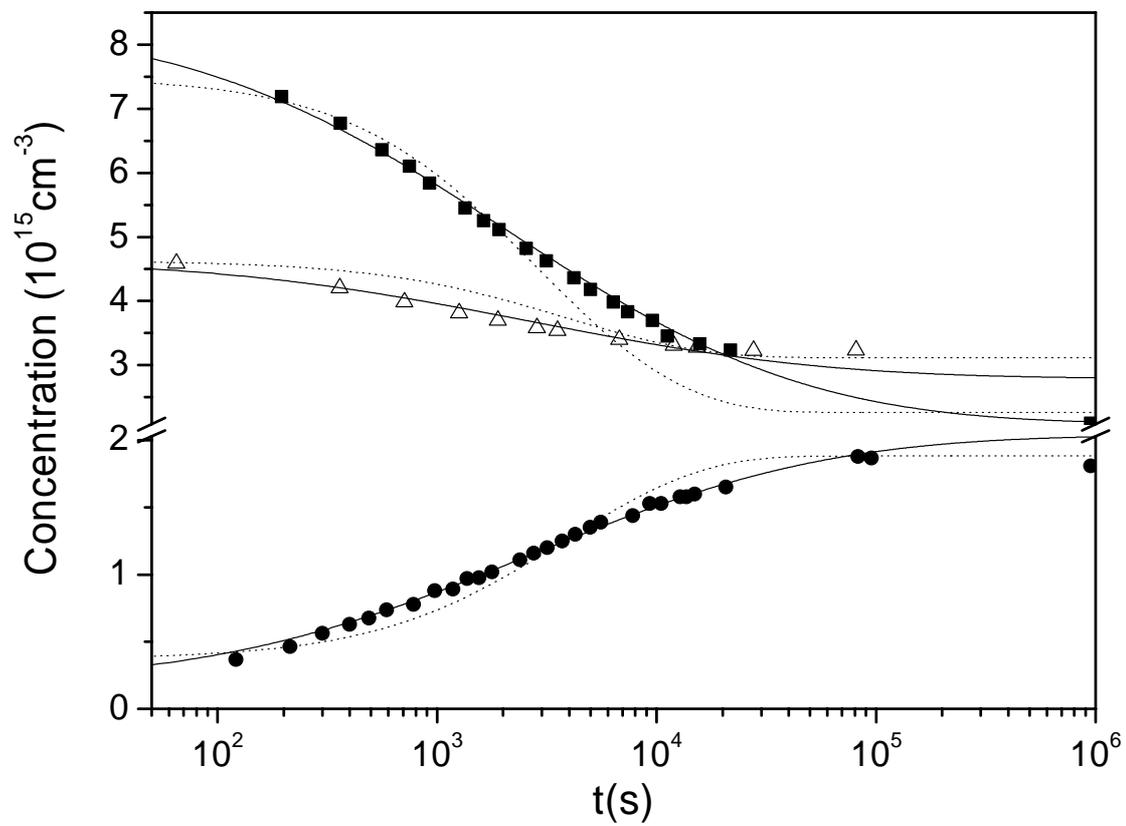

**FIGURE 7**